\documentclass[aps,prd,twocolumn,superscriptaddress,groupedaddress, showkeys]{revtex4} 
\usepackage{placeins}

\usepackage{subfig}
\usepackage{natbib}
\usepackage{orcidlink}
\usepackage{mathtools}
\usepackage{scalerel}
\usepackage{graphicx}  
\usepackage{bm}        
\usepackage{tikz}
\usepackage{amssymb}   
\usepackage{lipsum}
\usepackage{lineno}
\usetikzlibrary{svg.path}
\usepackage{amsmath,array}
\usepackage{hyperref}
\hypersetup{
	colorlinks=true,
	linkcolor=blue,
	filecolor=magneta,      
	urlcolor=blue,
}
\DeclareCaptionJustification{justified}{\leftskip=0pt \rightskip=0pt \parfillskip=0pt plus 1fil}

\definecolor{orcidlogocol}{HTML}{A6CE39}
\tikzset{
    orcidlogo/.pic={
        \fill[orcidlogocol] svg{M256,128c0,70.7-57.3,128-128,128C57.3,256,0,198.7,0,128C0,57.3,57.3,0,128,0C198.7,0,256,57.3,256,128z};
        \fill[white] svg{M86.3,186.2H70.9V79.1h15.4v48.4V186.2z}
        svg{M108.9,79.1h41.6c39.6,0,57,28.3,57,53.6c0,27.5-21.5,53.6-56.8,53.6h-41.8V79.1z M124.3,172.4h24.5c34.9,0,42.9-26.5,42.9-39.7c0-21.5-13.7-39.7-43.7-39.7h-23.7V172.4z}
        svg{M88.7,56.8c0,5.5-4.5,10.1-10.1,10.1c-5.6,0-10.1-4.6-10.1-10.1c0-5.6,4.5-10.1,10.1-10.1C84.2,46.7,88.7,51.3,88.7,56.8z};
    }
}
\newcommand\orcidicon[1]{\href{https://orcid.org/#1}{\mbox{\scalerel*{
                \begin{tikzpicture}[yscale=-1,transform shape]
                \pic{orcidlogo};
                \end{tikzpicture}
            }{|}}}}

\begin{document}

\preprint{APS/123-QED}

\title{Constraining Viscous Dark Matter in light of CMB Spectral Distortion}
\author{Shibsankar Si$^{\orcidicon{0009-0001-8038-976X}}$\,}
\email{p21ph001@nitm.ac.in}
\affiliation{National Institute of Technology, Meghalaya, Shillong, Meghalaya 793 003, India}
\author{Alekha C. Nayak$^{\orcidicon{0000-0001-6087-2490}}$\,}%
\email{alekhanayak@nitm.ac.in}
\affiliation{National Institute of Technology, Meghalaya, Shillong, Meghalaya 793 003, India}%
\author{Pravin Kumar Natwariya$^{\orcidicon{0000-0001-9072-8430}}$\,}
\email{pvn.sps@gmail.com}
\affiliation{School of Fundamental Physics and Mathematical Sciences, Hangzhou
Institute for Advanced Study, UCAS, Hangzhou, 310 024, China}
\affiliation{University of Chinese Academy of Sciences, Beijing, 100 190, China}

\date{\today}

\begin{abstract}
We calculate the $\mu$- and \textit{y}-type spectral distortions of Cosmic Microwave Background (CMB), taking a non-standard interaction between baryons and viscous dark matter. Using the CMB spectral distortion observations, we can constrain any exotic mechanism that may change the energy of the CMB photon, leading to a CMB spectrum distortion.  
Depending on the viscosity of dark matter, the energy transfer between dark matter and baryons may modify, leading to a modification in CMB  distortion. The existing Cosmic Background Explorer (COBE)/FIRAS and the Primordial Inflation Explorer (PIXIE) set limits on \textit{y} and $\mu$ types of distortions to $y = 1.5\times10^{-5}$, $\mu = 9.0\times10^{-5}$ and $y = 10^{-8}$, $\mu = 5.0\times10^{-8}$, respectively. In this paper, we discuss the pre-recombination contributions to $\mu$ and \textit{y}-type distortions by viscous dark matter and constrain the parameter space using PIXIE bounds on spectral distortion.
\end{abstract}

\maketitle


\section{\label{sec:level1} Introduction}

The Cosmic Microwave Background radiation represents the earliest light to have traversed the Universe. Its perfect blackbody spectra were initially discovered by the Cosmic Background Explorer (COBE) \cite{smoot1992structure, Mather:1993ij}. Subsequent generations of satellites, such as the WMAP and Planck satellites, have detected significant fluctuations in the power spectrum of CMB temperature on a large scale \cite{WMAP:2003zzr, WMAP:2012fli, Planck:2013oqw, Planck:2013pxb}. Additionally, ground-based observations, such as the Atacama Cosmology Telescope and the South Pole Telescope, have observed fluctuations on small scales \cite{Kosowsky:2003smi, SPT:2004qip}. These CMB fluctuations provide a window to explore the early Universe.

At redshift $z\gtrsim 2\times10^6$, any photon production or destruction in the Universe leads to a thermal equilibrium between matter and radiation through bremsstrahlung and double Compton scattering. However, as the Universe reaches redshift $z\lesssim 2\times10^6$, the efficiency of double Compton and bremsstrahlung scattering decreases because of the high entropy of our universe \cite{Chluba:2011hw, Zeldovich:1969ff, Sunyaev:1970er}. 
At redshifts between $5\times10^4$ and $2\times10^6$, any energy injection into the plasma causes the heated electrons to alter the photon temperature without changing the photon number density via Compton scattering. The equilibrium distribution of photons via Compton scattering after energy injection follows a Bose-Einstein distribution with a nonvanishing chemical potential, resulting in the production of $\mu$-type distortion \cite{khatri2012beyond}.
When energy injection occurs between redshifts $10^3$ and $5\times10^4$, the elastic Compton scattering of photons becomes inefficient in establishing the Bose-Einstein spectrum, leading to the production of $y$-type spectral distortion. After redshift $z\sim10^3$ (the recombination epoch), only $y$-type distortion from inverse Compton scattering of CMB photons by heated electrons in the intergalactic medium is expected. The COBE/FIRAS's constraints on the $y$-type ($\mu$-type) distortion are $y = 1.5\times10^{-5}$ $(\mu = 9.0 \times10^{-5})$ \cite{Fixsen:1996nj}. The future experiment Primordial Inflation Explorer (PIXIE) aims to detect $y$-type ($\mu$-type) distortion at levels of $y = 10^{-8}$ ( $\mu= 5.0\times10^{-8}$) at a $5\sigma$ level \cite{Kogut:2011xw}. In this paper, we use these limits on spectral distortion to constrain the parameter space of viscous dark matter mass and the scattering cross-section with baryons.

Distortions of the CMB spectrum may occur from any mechanism that can heat or cool the baryons. Processes such as dark matter annihilation and decay can inject energy into baryonic matter, resulting in noticeable CMB distortions \cite{Chluba:2011hw}. Therefore, the study of CMB distortion can give a strong upper limit on the dark matter annihilation rate. When unstable particles decay at a rate greater than the Hubble rate, a large quantity of energy can be injected into the CMB. In the context of CMB spectral distortion, the constraints on the decay rate of unstable particles have been studied in the earlier articles \cite{Chluba:2011hw, Hu:1993gc}. Furthermore, the decaying primordial magnetic fields (PMFs) can also inject energy into the plasma via ambipolar diffusion and turbulent decay \cite{Sethi:2004pe, Tashiro:2005ua, Tashiro:2006uv, Schleicher:2008hc, Bhatt2019pac, Minoda:2018gxj, Natwariya:2020,  Natwariya:2021, Natwariya:2023T}. The energy injection by decaying PMFs can also lead to CMB spectral distortion \cite{Shaw:2010ea}. In Ref.\cite{Kunze:2013uja}, the authors calculated the $\mu$- and $y$-types of spectral distortion and constrained on the PMFs strength, $B_0$, and the spectral index of PMFs, $n_B$. Similarly, primordial black holes (PBHs) can also inject energy into CMB by Hawking evaporation. PBHs with masses less than $\sim10^{15}$~g may have evaporated by now, and their high photon production during evaporation makes them significant contributors to CMB distortions. The effects of PBHs evaporation on CMB distortions have been studied in Ref. \cite{Tashiro:2008sf}, where it was shown that PBHs with masses in the range from $10^{11}$~g to $10^{14}$~g can significantly impact CMB spectral distortions. Additionally, primordial black holes (PBHs) are important sources of energy injection into the CMB. PBHs having masses smaller than $10^{15}$~g have evaporated from the Universe, and their high photon production during evaporation makes them significant contributors to CMB distortions \cite{Tashiro:2008sf}.

 Despite a decade of research, understanding the microscopic nature of dark matter is still one of the main challenges and active areas of research in cosmology and astroparticle physics. The standard model of cosmology, $\Lambda$CDM model, posits that the Universe is predominantly composed of collisionless cold dark matter (about $27$ percent of the total energy budget of the Universe) particles that only interact with standard model (SM) particles through gravity. In this framework, dark matter is considered an ideal fluid. This model effectively explains the large-scale structure of the universe. However, over time, many discrepancies between the observations and predictions based on the $\Lambda$CDM framework have emerged, such as the missing satellites problem, too-big-to-fail problem, core-cusp problem,  etc. (for details, see the review article \cite{Bullock:2017} and references therein). There have been numerous efforts to resolve these small-scale problems, and over time, many alternatives to the cold dark matter model have been proposed, such as self-interacting dark matter \cite{Spergel:2000DSP, Tulin:2017ara, Kaplinghat:2016TY}, fuzzy cold dark matter \cite{Hu:2000, Schive:2014, purohit2021rotating}, primordial black holes as dark matter \cite{Bird:2016, Abbott:2016A, Abbott:2016B, Abbott:2016C, Abbott:2016D, Zeldovich:1967, Hawking:1971, Carr:1974H, Carr:1975, Natwariya:2021PBH, Sasaki:2016}, warm dark matter \cite{Blumenthal:1982, Dodelson:1994, Colombi:1996, Sitwell:2014, Brdar:2018, Natwariya:2022}, etc. One proposed approach to addressing these issues involves considering dark matter particles as warm rather than cold, suggesting that they were quasi-relativistic at the time of kinetic decoupling from the thermal bath in the early Universe \cite{Tulin:2017ara}. In contrast to CDM, warm DM predicts a power spectrum that is damped on small scales due to free-streaming, resulting in a reduction in the number of substructures. Warm DM halos form later than CDM halos, leading to lower concentration levels in warm DM halos. High-resolution N-body simulations have indicated that warm DM can offer solutions to the problems of missing satellites and too big-to-fail \cite{Tulin:2017ara}. Furthermore, warm DM halos are characterised by constant-density cores determined by the phase space density limit. However, these core sizes are too small to resolve the core-cusp problem under Lyman-$\alpha$ forest observations \cite{ Macci__2012, Carucci:2015bra, Tulin:2017ara}.

 Spergel and Steinhardt, in 1999, proposed a solution to the core-cusp and missing satellites problem by introducing the existence of self-interacting dark matter (SIDM) \cite{Spergel:1999mh}. In this scenario, DM particles interact with each other elastically. This can give rise to a non-zero viscosity in DM fluid. The viscosity should be considered when looking at departures from thermal equilibrium to the first order in the cosmic fluid. The viscosity of SIDM can be characterised by two separate coefficients: bulk viscosity and shear viscosity. Here, it is to be noted that the viscosity in dark matter can also arise in the $\Lambda$CDM description as we can take fluid approximation for dark matter on the nonlinear scale because the particles can move freely up to a finite scale under gravity \cite{Natwariya:2020V}. However, in the present work, we consider a model-dependent bulk viscosity for self-interacting dark matter. The widely accepted spatial isotropy of the Universe leads to the omission of shear viscosity in most analyses \cite{Floerchinger:2014jsa}. The bulk viscosity can play a crucial role in cosmological models. If the bulk viscosity is sufficiently large, it can provide an explanation for dark energy \cite{Sasidharan_2015, Arora_2022}. It offers a compelling explanation for the late-time accelerated expansion of the Universe \cite{Sasidharan_2015, Arora_2022}. The introduction of bulk viscosity for cosmic fluid can also circumvent the Hubble tension \cite{Normann_2021}. Studies have demonstrated that bulk viscosity can be generated by the decay of cold dark matter into relativistic particles \cite{Brevik:2017msy, Anand:2017ktp}. Furthermore, it has been suggested that dark matter may exhibit inflation-like behaviour if it possesses a high enough bulk viscosity \cite{Padmanabhan:1987dg, Gron:1990ew, Brevik:2017msy, Anand:2017ktp}. These results show the potential impact of DM viscosity on the evolution of the Universe and its implications for cosmological models.

  The temperature of dark matter rises if the dark matter's viscosity is high enough \cite{Weinberg:1972kfs}. However, the low DM viscosity does not generate sufficient entropy to change the DM temperature, and the DM fluid behaves like an ideal fluid. Using the law of thermodynamics,  the temperature evolution of baryon and dark matter in the presence of DM-gas interaction has been studied \cite{Bhatt:2019qbq}. 
The energy transfer from baryon to dark matter changes due to the interaction of viscous dark matter with baryon.
 Hence, the photon energy also gets modified as it is coupled with the baryon during the pre-recombination, and it leaves an imprint on the CMB spectrum. Using PIXIE and FIRAS limits on CMB spectral distortion, we constrain the parameter space of viscous dark matter.

 This paper is organised as follows: In Sec.\ref{sec2}, we discuss the evolution of dark matter temperature in the presence of viscosity. In Sec.\ref{sec3}, we briefly
discuss the temperature evolution of baryon and dark matter in the presence of baryon-dark matter cross-section and dark matter viscosity. In Sec.\ref{sec4}, we briefly review the CMB spectral distortion. The evolution of  CMB spectral distortion for different viscosity parameters, dark matter mass, and baryon-dark matter cross-section are presented in Sec.\ref{sec5}. Finally, we put constraints on dark matter mass and dark matter baryon scatter cross section using PIXIE and FIRAS limit in Sec.\ref{sec6}.

\section{VISCOUS DARK MATTER}
\label{sec2}
In the standard $\Lambda$CDM  cosmological model, dark matter is taken to be collisionless and pressureless ideal fluid. It fits data very well on a large scale; however, on a small scale, it can not explain the core-cusp problem, missing satellites problem, and too-big-to-fail problem. Collisionless and pressureless cold dark matter lead to excess structure and clustering. These clustering problems can be solved by taking self-interaction of dark matter. Self-interaction of dark matter results in an effective bulk viscosity. Using the Eckart theory of dissipative process, the phenomenological form of dark matter viscosity can written as \cite{Velten:2013pra}, 
\begin{eqnarray}
     \zeta_{\chi} = \zeta_{0}\left( \frac{\rho_{\chi}(z)}{\rho_{\chi_0}}\right)^{\gamma}~~,
\end{eqnarray}
Here $\rho_{\chi}(z)$ and $\rho_{\chi_0}$ represent dark matter energy density at redshifts $z$ and today, respectively. $\gamma$ is the viscosity parameter, we take  $\gamma=0, ~-1/2$ so that the integrated Sachs-Wolfe (ISW) effects problem for these viscosity models is less severe \cite{Velten:2011bg, Velten:2012uv}. $\zeta_{0}$ is another viscosity parameter which is related to another dimensionless viscosity parameter: $\bar{\zeta}={24\, \pi\, G\, \zeta_{0}}/H_0$, where $G$ is gravitational constant and $H_0$ is Hubble constant.

The $\rho_{\chi}(z)$ we can calculate by using the continuity equation  \cite{Velten:2013pra}.
\begin{eqnarray}
\frac{d\Omega_{\chi}(z)}{dz} &= \frac{3}{1+z}\Omega_{\chi}(z)  - \frac{\bar{\zeta}}{1+z}\left( \frac{\Omega_{\chi} (z)}{\Omega_{\chi}(0)}\right)^{\gamma}\big[ \Omega_{r0}(1+z)^4 \nonumber \\
&+ \Omega_{b0}(1+z)^3 + \Omega_{\mathrm{\chi}}(z) + \Omega_{\Lambda}\big]^{1/2} = 0\,,\label{eq:viscevol}
\end{eqnarray}
 with initial condition $\Omega_{\chi}(0)=\Omega_{\chi_{0}}$, where $ \Omega_{\mathrm{i}}(z) = {4\,\pi\, G\, \rho_{\mathrm{i}}(z)}/{3\,H^{2}_{0}} $, where $ i=r,\ b,\ \chi,\ \Lambda $. Here, the 0 subscript represents the present relic density of different species at the present time. In the presence of  viscous dark matter, the Hubble expansion rate is
\begin{equation}
H = H_{0}\bigg[ \Omega_{r0}(1+z)^4 + \Omega_{b0}(1+z)^3 + \Omega_{\chi}(z) + \Omega_{\Lambda} \bigg]^{1/2}\,. \label{eq:hubble}
\end{equation}
 
Viscosity in the dark matter results in entropy production, which heats the dark matter. Using the laws of thermodynamics, the temperature evolution of dark matter can be written as 
\begin{eqnarray}\label{23B}
     (1+z)\frac{dT_{\mathrm{\chi}}}{dz} = 2T_{\chi} - \frac{2m_{\chi}}{3H\rho_{\chi}}\frac{dQ_{\mathrm{v}}}{dVdt}\,,
\end{eqnarray}
here,  $T_\chi$, $m_\chi$, $\rho_\chi$, and $H(z)$ are the DM temperature, mass, energy density, and the Hubble parameter, respectively. In the above equation, the first term represents the decrease in dark matter temperature due to Hubble expansion of the Universe, and the second term represents the heating of dark matter due to dark matter viscosity. 

The entropy production per unit volume due to the bulk viscosity of dark matter in the expanding Universe can be written as  \cite{Bhatt:2019qbq},
\begin{eqnarray}
    {\nabla_{\mu}S^{\mu} = \frac{\zeta_{\chi}}{T_{\chi}}\left( \nabla_{\mu}u^{\mu}\right)^{2} }\,,
\end{eqnarray}
where, entropy four-vector $S^{\mu}=n_{\chi} s_{\chi}u^{\mu}$, 
 $s_{\chi} $ is the entropy per unit particle,  $n_{\chi}$ is the  number density, $T_{\chi}$ is the temperature and $u^{\mu}$ is the four-velocity of the dark matter respectively. Applying the second law of thermodynamics, the heat energy produced per unit volume per unit of time due to dark matter viscosity in the comoving frame is given by  \cite{Bhatt:2019qbq}
 \begin{eqnarray}
     \frac{dQ_{\mathrm{v}}}{dVdt} =   T_{\chi}\nabla_{\mu}S^{\mu}= \zeta_{\chi}\left( \nabla_{\mu}u^{\mu}\right)^{2} =  \zeta_{\chi}(3H)^{2}
     \label{23A}
 \end{eqnarray}
From Eq.\eqref{23B} and (\eqref{23A}), we get the temperature evolution equation as
\begin{eqnarray}\label{vtemp}
     (1+z)\frac{dT_{\mathrm{\chi}}}{dz} = 2T_{\chi} - \frac{6m_{\chi}\zeta_{\chi}H}{\rho_{\chi}}\,.
\end{eqnarray}
Eq.\eqref{vtemp} implies that the temperature of dark matter rises if the dark matter's viscosity is high enough.

\section{Evolution of the IGM in light of viscous dark matter}
\label{sec3}

In this section, we discuss how the viscous property of dark matter affects the evolution of dark matter and IGM gas temperature. In the standard $\Lambda$CDM, dark matter is much colder than the baryon as dark matter decoupled very early from the plasma. Hence, the interaction between baryon and dark matter results in lower baryon temperature \cite{Tashiro_2014, Mu_oz_2015, barkana2018possible}. Interaction between dark matter and baryon is taken in the form of  $\sigma=\sigma_0 v_{rel}^{-4}$, where $v_{rel}$ is the relative velocity between dark matter and gas, In the presence of baryon-dark matter interaction,  the heating rate of baryon can be expressed as \cite{Tashiro_2014, Mu_oz_2015, barkana2018possible} 
 \begin{eqnarray}\label{21}
   \frac{dQ_b}{dt}=\frac{2m_b\rho_\chi\sigma_0e^{-\frac{r^2}{2}}(T_\chi-T_b)}{(m_b+m_\chi)^2\sqrt{2\pi}u^3}+\frac{\rho_\chi m_\chi m_b v_{rel} D(v_{rel})}{\rho_m(m_\chi+m_b)}\,,\label{dqbdt}
 \end{eqnarray}
 where, $T_b$  represents baryon temperature, $m_b$ mass of the baryon, $r={v_{rel}}/{u}$, and $u^2 = \big({T_b}/{m_b} + {T_{\chi}}/{m_{\chi}}\big)$. The second term in Eq.\eqref{21} represents the heating of both baryon and dark matter due to drag between baryon and dark matter fluid. The drag term is expressed as \cite{Mu_oz_2015},
\begin{eqnarray}
D(v_{rel}) \equiv - \frac{dv_{rel}}{dt} = \frac{\rho_m\sigma_0}{m_b + m_\chi}\frac{1}{(v_{rel})^2}F(r)\,,
\end{eqnarray}
where. the function $F(r)$ is given by 
\begin{eqnarray}
F(r) = \mathrm{erf} \left(\frac{r}{\sqrt{2}}\right) - \sqrt{\frac{2}{\pi}}e^{-\frac{r^2}{2}}r\,.
\end{eqnarray}
In the pre-recombination era, where the estimated rms value of relative velocity between DM and baryon ($v_{rms}$) is $10^{-4} c$ \cite{Slatyer:2018aqg}. Hence,  second term of Eq.\eqref{21} do not contribute the heat rate of the baryon  $D(v_{rel}) $ vanishes in the pre-recombination era 

We can get the dark matter heating rate (${dQ_\chi}/{dt}$) by exchanging $b\leftrightarrow x$ in Eq.\eqref{21}.
Now, the temperature evolution equations of the baryon and dark matter will be\cite{Bhatt:2019qbq} 
\begin{eqnarray}\label{23}
(z+1)\frac{dT_{b}}{dz}={2T_{b}}-\frac{\Gamma_C}{H}(T_{\gamma}-T_{b})-\frac{2}{3H}\frac{dQ_{\mathrm{b}}}{dt}\,,\label{dtbdz}
\end{eqnarray}
\begin{eqnarray}\label{24}
    (1+z)\frac{dT_{\mathrm{\chi}}}{dz} = 2T_{\chi} - \frac{6m_{\chi}\zeta_{\chi}H}{\rho_{\chi}} - \frac{2}{3H}\frac{dQ_{\chi}}{dt}\,,\label{dtxdz}
\end{eqnarray}
where, $\Gamma _C$ denotes the Compton scattering rate. In Eq.\eqref{23}, the last term ${dQ_{\mathrm{b}}}/{dt}$ represents the rate of heat transfer from baryon to dark matter caused by the interaction of DM and baryon. Further, in Eq.\eqref{24}, the second term ${dQ_{\chi}}/{dt}$ denotes the dark matter particle's heat absorption rate. Dark matter is cooling due to Hubble expansion, which is the first term in Eq.\eqref{24}, the second term represents the heating of dark matter due to the viscous nature of dark matter, and the third term represents the heating of dark matter due to heat transfer from baryon.

 \section{CMB spectral distortions}
\label{sec4}
The measurement of the CMB spectral distortions is a powerful tool for studying the evolution of the Universe. Two spectral distortions commonly identified for CMB are $\mu$-type and \textit{y}-type distortions. The $\mu$-type distortion arises when the number density of photons remains the same but the total energy density increases. This effect can result in an effective non-zero chemical potential and establish a Bose-Einstein distribution for the photons. This type of distortion arises at a very high redshift, $z\gtrsim5\times10^4$.
 
On the other hand, the \textit{y}-type distortion is produced relatively at lower redshift when Compton scattering becomes insufficient. Here, non-relativistic Compton scattering becomes important, giving rise to y-type distortion. Both types of distortion, in the light of non-standard physics, have been studied in various prior studies \cite{Chluba:2011hw, Sunyaev:1970er, Zeldovich:1969ff, khatri2012beyond}. The COBE/FIRAS put an upper bound on the CMB spectral distortions \cite{Fixsen:1996nj}. In this work, we consider the upper bound on CMB spectral distortion by PIXIE and COBE/FIRAS, and constrain the parameter space for viscous dark matter.

 At redshift $z\gtrsim 2\times10^6$, photons remain in complete thermal equilibrium with the plasma due to efficient scatterings between photons and baryons. Any energy injection into the medium is quickly redistributed in photons by Compton scattering, and the modification in the number density of photons is adjusted by double Compton and bremsstrahlung, keeping a black body distribution for photons.
 Below $z\lesssim 2\times10^6$, bremsstrahlung and double Compton scattering become ineffective due to the high entropy of the Universe. Between redshift $5\times10^4 \lesssim z\lesssim 2\times10^6 $,  any energy injection into the IGM changes the photon energy density while maintaining the photon number density constant through elastic Compton scattering,  resulting in  Bose-Einstein distribution. This leads to an effective non-zero chemical potential, known as $\mu$-type distortion \cite{khatri2012beyond, Kunze:2013uja}.

 The evolution of $\mu$ can be  expressed as \cite{Kunze:2013uja}
 \begin{eqnarray}\label{dmu/dt}
     \frac{\partial \mu}{\partial t}= \frac{1.4}{3}\left(\frac{dQ_b}{dt}\Big/\rho_\gamma\right) -\frac{\mu}{t_{\rm dc}}
 \end{eqnarray}
In the first term, a factor of $1/3$ denotes that spectral distortions arise by only $1/3$ of the total energy injection, while the rest $2/3$ contributes to increasing the average temperature \cite{Chluba:2012gq, Pajer:2012qep, Kunze:2013uja}. Here, the double Compton scattering time scale is denoted by $t_{\rm dc}$
\begin{eqnarray}
    t_{\rm dc} =2.06\times 10^{33}\left(\frac{1}{\Omega_b h^2}\right)\left(1-\frac{Y_p}{2}\right)^{-1}z^{-9/2}
\end{eqnarray}
where, $\Omega_b$ is the dimensionless energy density parameter for the baryons, and $Y_p$ is the primordial abundance of helium, which is estimated to make up roughly $25\%$ of the Universe's ordinary matter. The value of the primordial helium abundance $Y_p\approx 0.24$. At redshift between $2\times 10^6$ to $5\times 10^4$ the solution of the Eq.\eqref{dmu/dt} can be written as  \cite{Kunze:2013uja}

 \begin{eqnarray}
     \mu = \frac{14}{30}\int  e^{-\left({z}/{z_{\rm dc}}\right)^{5/2}}\ \left[\left({\frac{dQ_b}{dz}}\right)\Big/{\rho_\gamma}\right]\ dz\,,\label{dmudz}
 \end{eqnarray}
here, the integration limits are from $z=2\times 10^6$ to $z=5\times 10^4$. Furthermore, $dQ_b/dz$ represents the energy injection with redshift, $\rho_\gamma$ is the photon energy density and 
\begin{eqnarray}
z_{\rm dc}\equiv 1.1 1\times 10^7\left(\frac{\Omega_bh^2}{0.0224}\right)^{-{2}/{5}}\,,
\end{eqnarray}
here, $\Omega_b$ is the dimensionless energy density parameter for the baryons.


 The $\mu$ type of spectral distortion is defined by the frequency-dependent chemical potential, which is almost constant at the high frequency and vanishes at the low frequency. 
 Compton scattering is no longer effective for redshifts $z\leq 5\times 10^4$ because the energy photon decreases to a sufficiently low value.
 Hence, the Compton redistribution of photons over frequencies is too weak to establish the Bose-Einstein spectrum. Any energy injection into the IGM  no longer produces $\mu$-type spectral distortion \cite{khatri2012beyond}. Between $10^3 \lesssim z\lesssim 5\times10^4 $, any energy injection into IGM produces \textit{y} type of distortion, where low-energy CMB photons scatter from high-energy cluster electrons, and photons gain energy by interacting with electrons via the inverse Compton scattering \cite{Sunyaev:1970er}. The CMB spectrum shifts to a higher frequency due to the energy transfer from the hotter electron to a colder photon, and it produces \textit{y}-type spectral distortion. 
The CMB spectral distortion due to inverse Compton scattering is defined by Comptonisation parameter \textit{y} \cite{Kunze:2013uja, khatri2012beyond}. The \textit{y} parameter is given by,
\begin{eqnarray}
y = \frac{1}{12} \int \ \left[\left({\frac{dQ_b}{dz}}\right)\Big/{\rho_\gamma}\right]\ dz\,,
 \end{eqnarray}
here, the integration limits are from  $z=5\times 10^4$ to the recombination epoch, $z \sim 10^3$. After the recombination epoch, photons and baryons get decoupled. However, the free electrons and photons remain in thermal equilibrium due to Compton scattering till around a redshift $z \approx 200$. Only \textit{y}-type distortion was observed in this era, which we have not studied in this paper.

In this paper, we have taken the effect of dark matter viscosity during the pre-recombination era. If the DM viscosity is sufficient, the dark matter temperature rises over cosmic time due to entropy formation. Since dark matter is interacting with the baryon, it also affects the baryon temperature. Taking both baryon-dark matter interaction and viscosity of dark matter,  we calculate the CMB spectral distortion for different viscosity parameters and DM-baryons scattering cross-section. We also constrained DM parameter space from the spectral distortion.

\section{Discussion and Results}
\label{sec5}
 This work discusses the $\mu$ and \textit{y} type of CMB spectral distortion in light of dark matter viscosity. In this section, we plot variations of $\mu$-type and $y$-type distortion as a function of redshift for different dark matter mass and baryon-dark matter scattering cross-sections. Furthermore, we also plot the allowed parameter space for dark matter mass and cross-section with baryons using PIXIE limits on $\mu$ and $y$ types of CMB spectral distortions. 

Here, we will study how the interactions between DM and baryons change their temperature. To do so, we need to calculate the drag on the relative velocity due to interactions with baryons \cite{Munoz:2015bca}. We consider that the DM elastically scatters off baryons with a scattering cross-section $\sigma=\sigma_0v_{rel}^n$, where $v_{rel}$ is the magnitude of the baryon-DM relative velocity and power index $n$ is an integer \cite{Munoz:2015bca}. At a very early time, baryons always have a single temperature because Coulomb scatterings occur frequently. The CMB and baryon temperatures are closely coupled due to the Compton scattering with free electrons and photons. The DM temperature's evolution and the time it deviates from the baryon temperature depend on the power index $n$ \cite{Slatyer:2018aqg}. Baryons and dark matter are coupled early for larger $n$; however, as the Universe expands, the DM cools adiabatically as the scattering rate falls below the Hubble rate \cite{Ali-Haimoud:2015pwa, Slatyer:2018aqg}. Here, we have considered the scale factor $a_{\chi b}$ as \cite{Ali-Haimoud:2015pwa}

\begin{eqnarray}
    a_{\chi b}=\left( \frac{2c_n\sigma_0\rho_c\Omega_{b}\left(\frac{T_\gamma^0}{m_b}\right)^{\frac{n+1}{2}}}{H_0\Omega_{r}^{1/2}} \frac{m_b}{m_\chi}\left( 1+\frac{m_b}{m_\chi} \right)^{\frac{n-3}{2}}\right)^{\frac{2}{n+3}}\label{axb}
\end{eqnarray}
where $\rho_c$, $\Omega_{b}$, and $\Omega_{r}$ are the critical density, the baryon density, and radiation density, respectively. Here $c_n=\frac{2^\frac{n+5}{2}\Gamma\left(3+\frac{n}{2}\right)}{3\sqrt{\pi}}$ \cite{Slatyer:2018aqg}. The transition between thermal coupling and decoupling of DM is shown by the scale factor $a_{\chi b}$. The DM initially becomes strongly coupled to the baryons and thermally decouples at scale $a>a_{\chi b}$. In the range of DM masses that we look at, DM temperature becomes very low at redshift $z<10^{6}$ ($10^{-6}$ times lower than the baryons temperature)\cite{Slatyer:2018aqg}. Here we consider the pre-recombination era (redshift $z>10^3$), where the estimated rms value of relative velocity between DM and baryon ($v_{rms}$) is $10^{-4} c$ \cite{Slatyer:2018aqg}, and the free electron fraction $x_e\approx1$.

\begin{figure*}
    \begin{center}
        \subfloat[] {\includegraphics[width=3.5in,height=2.5in]{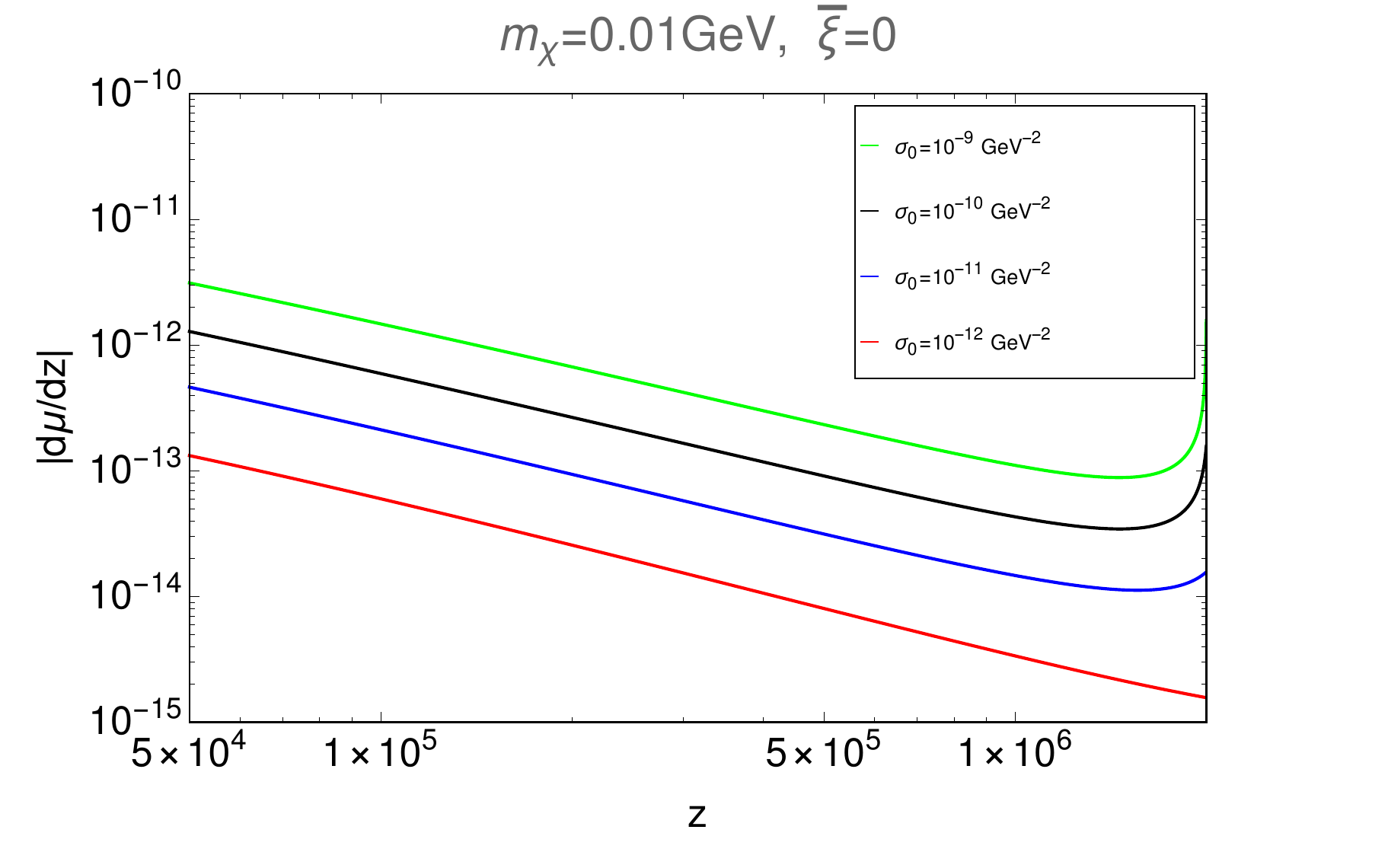}\label{p1a}}
        \subfloat[] {\includegraphics[width=3.5in,height=2.5in]{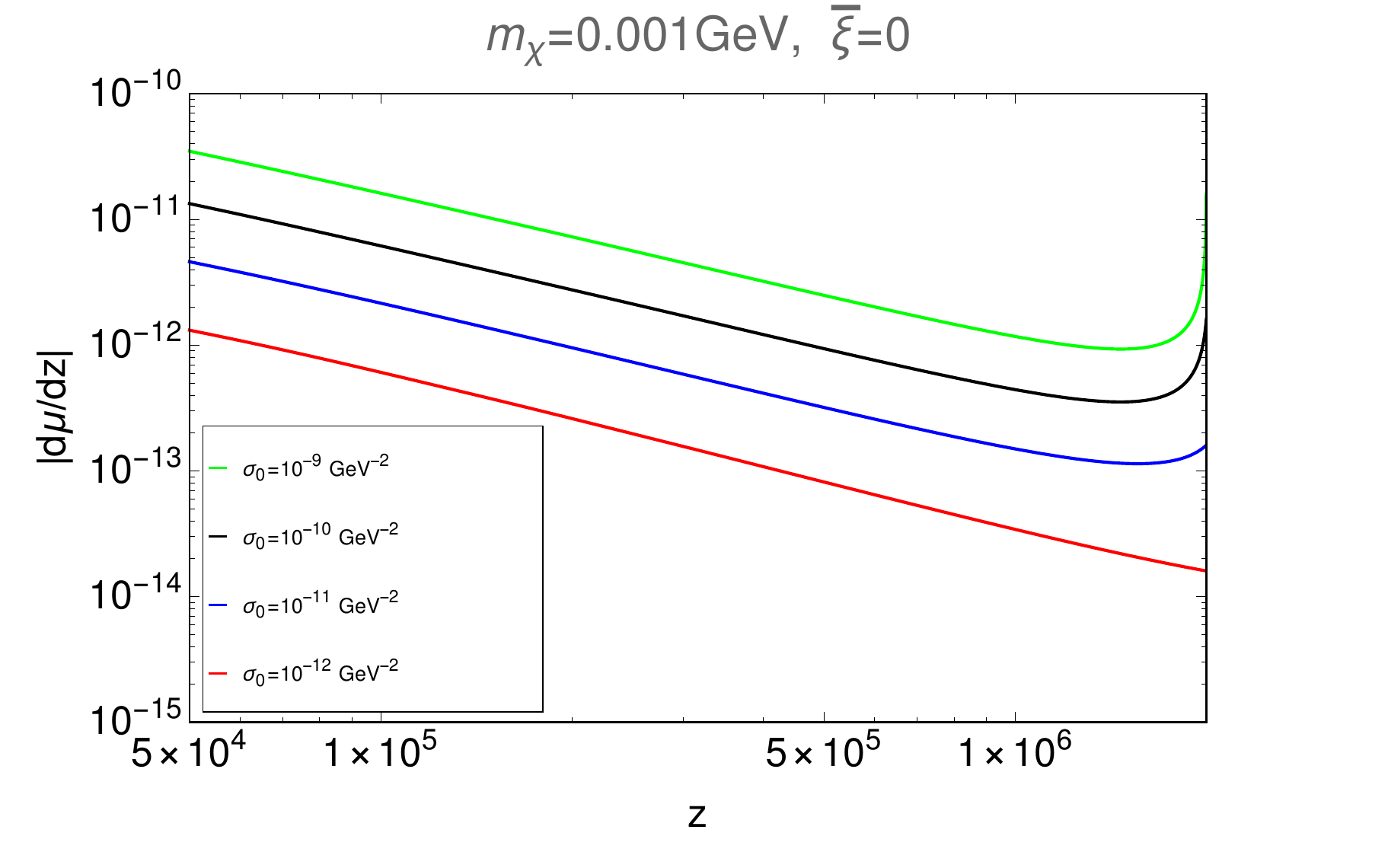}\label{p1b}} \\
        \subfloat[] {\includegraphics[width=3.5in,height=2.5in]{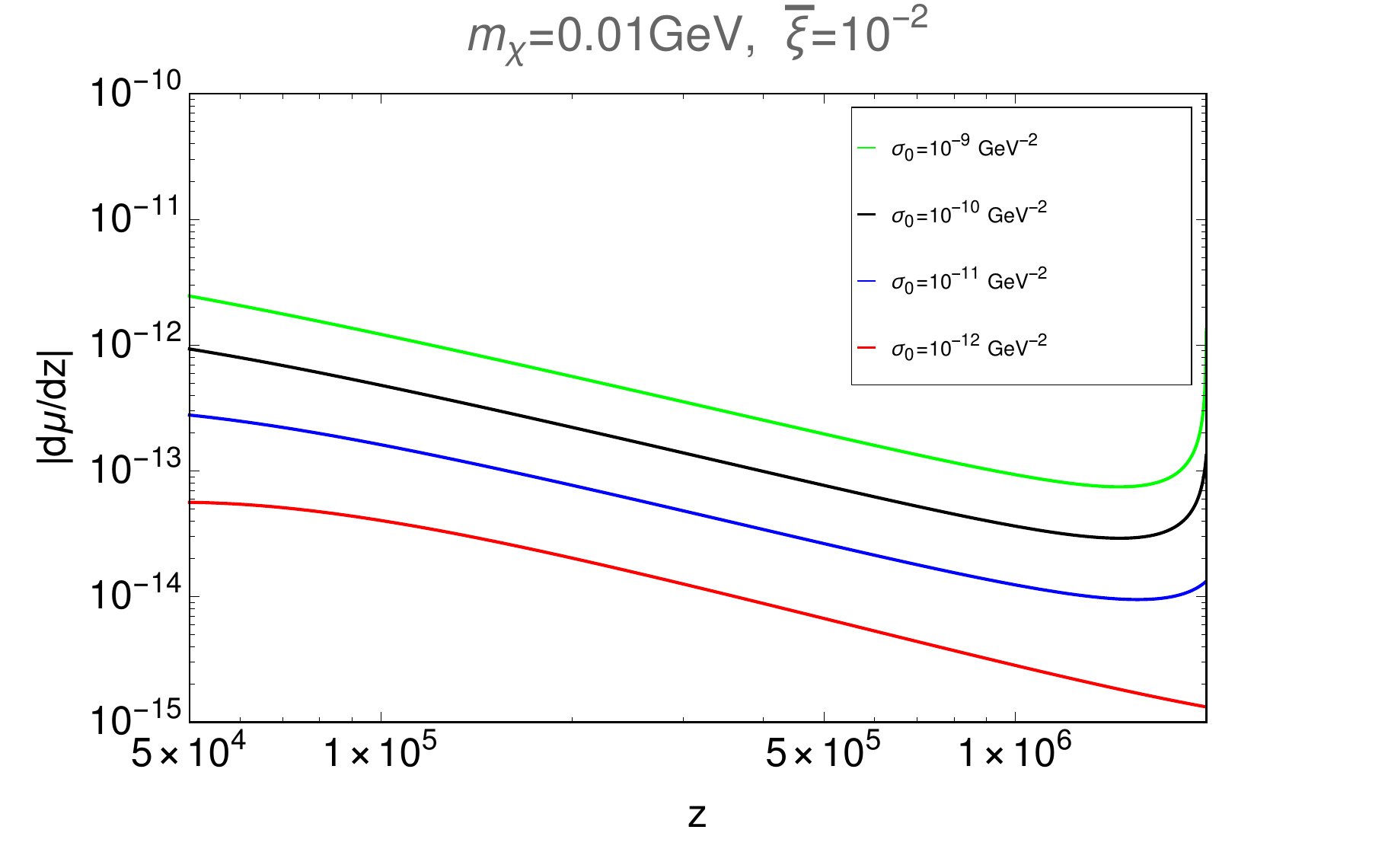}\label{p1c}}
        \subfloat[] {\includegraphics[width=3.5in,height=2.5in]{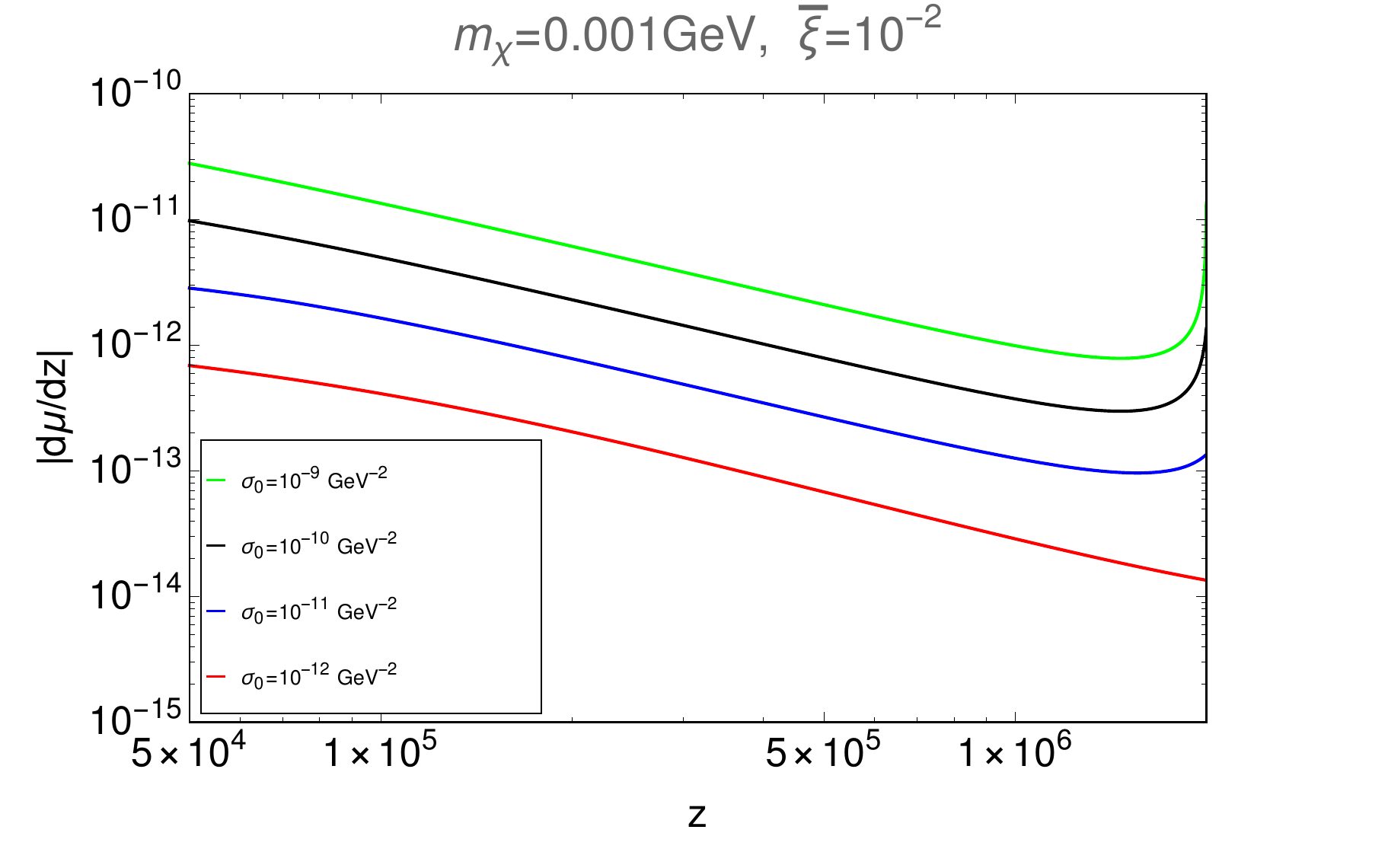}\label{p1d}}
    \end{center}
    \caption{Evolution of ${d\mu}/{dz}$ as a function of redshift. Here, we vary the strength of the dark matter cross-section with baryons as $\sigma_0= 10^{-12},\ 10^{-11},\ 10^{-10},\ 10^{-9}$~GeV$^{-2}$, for fixed values of the viscosity parameters ($\gamma$ and $\zeta_0$) and mass of dark matter. In the left panels, dark matter mass is fixed to $m_{\chi}=0.01$~GeV, while in the right panels, it is fixed to $m_{\chi}=0.001$~GeV. In the figure, we have varied the value of $\zeta_0$ from $0$ to $10^{-2}$ from top to bottom.}\label{plot:1}
\end{figure*}

\begin{figure*}
    \begin{center}
        \subfloat[] {\includegraphics[width=3.5in,height=2.5in]{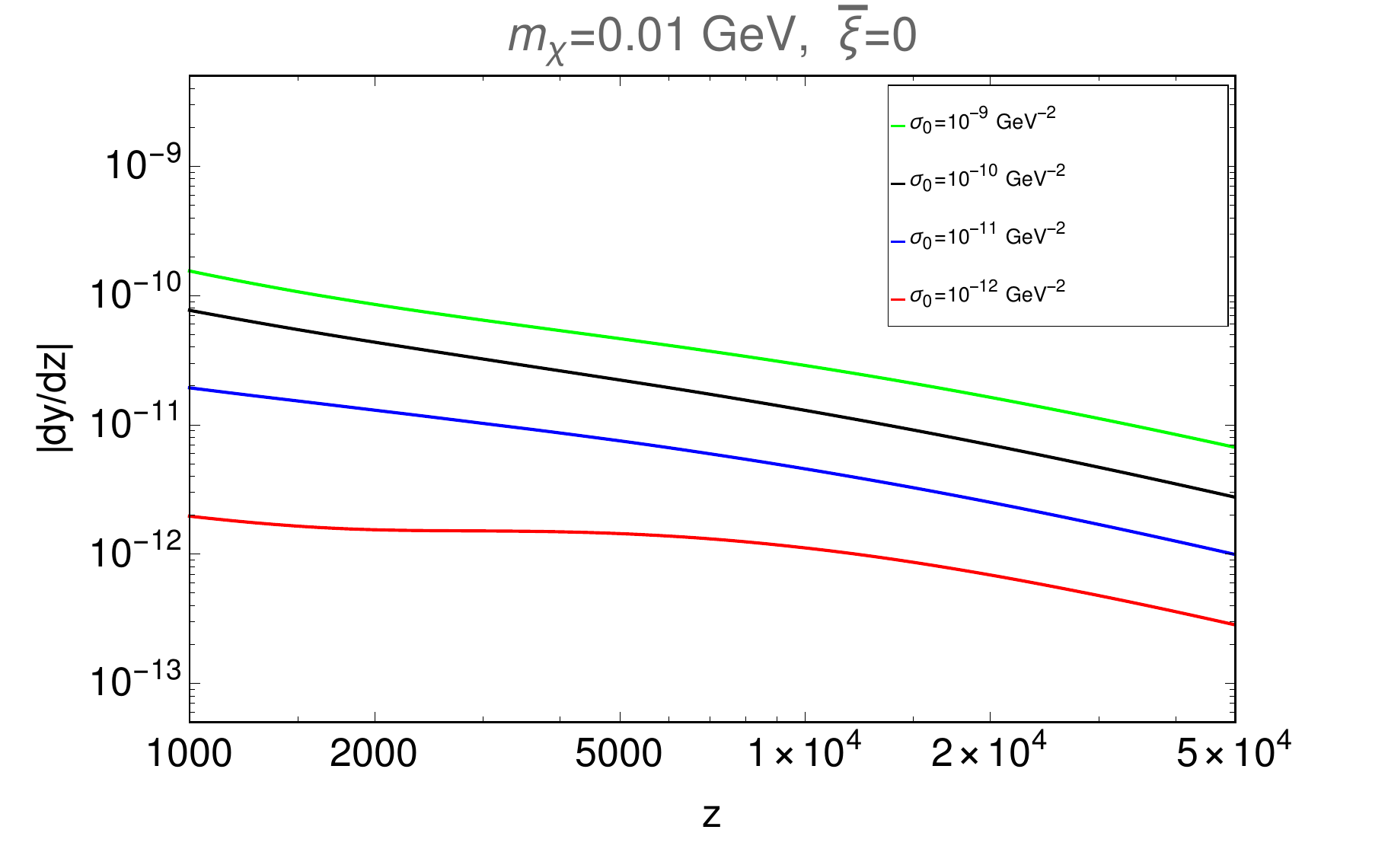}\label{p2a}}
        \subfloat[] {\includegraphics[width=3.5in,height=2.5in]{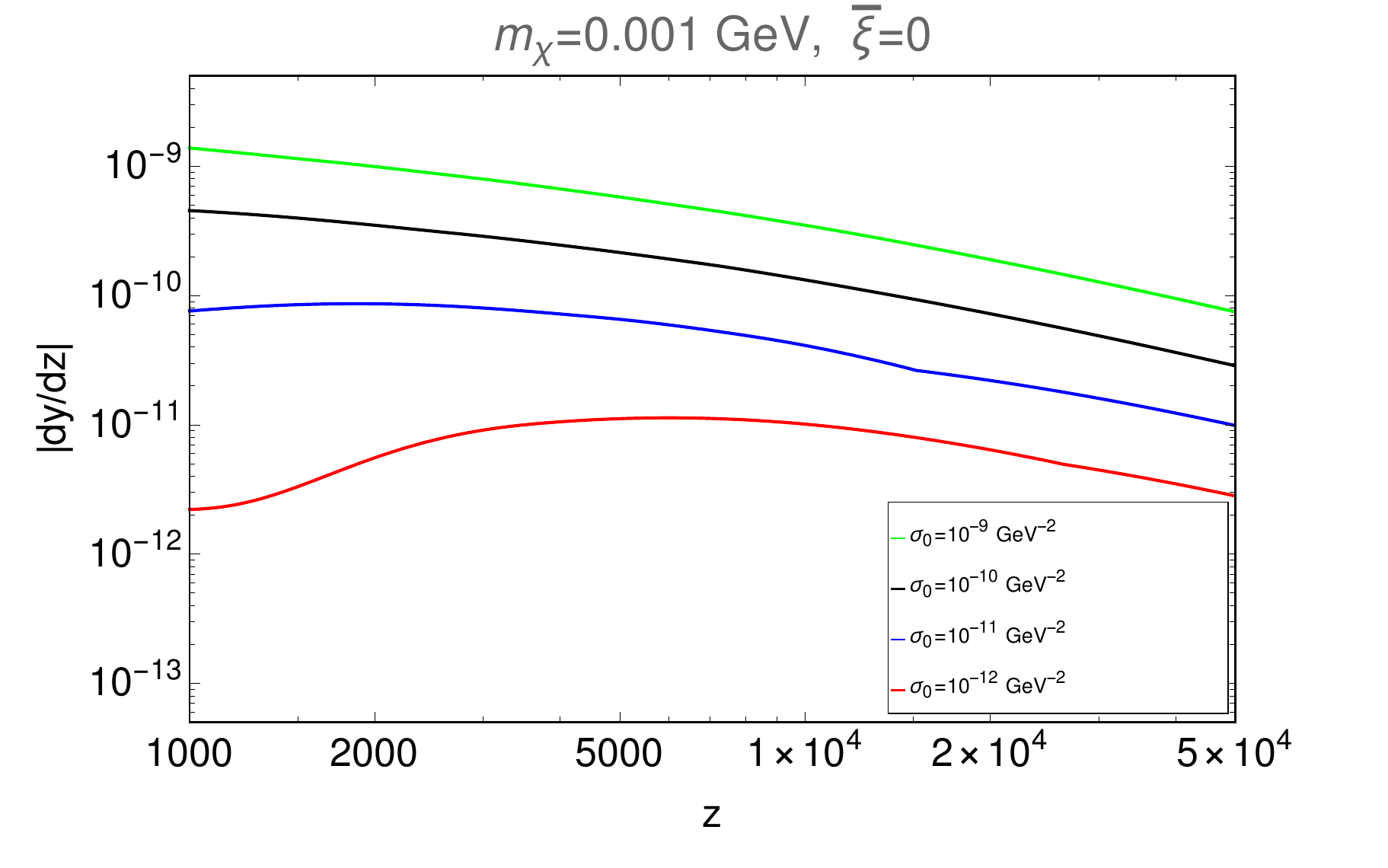}\label{p2b}} \\
        \subfloat[] {\includegraphics[width=3.5in,height=2.5in]{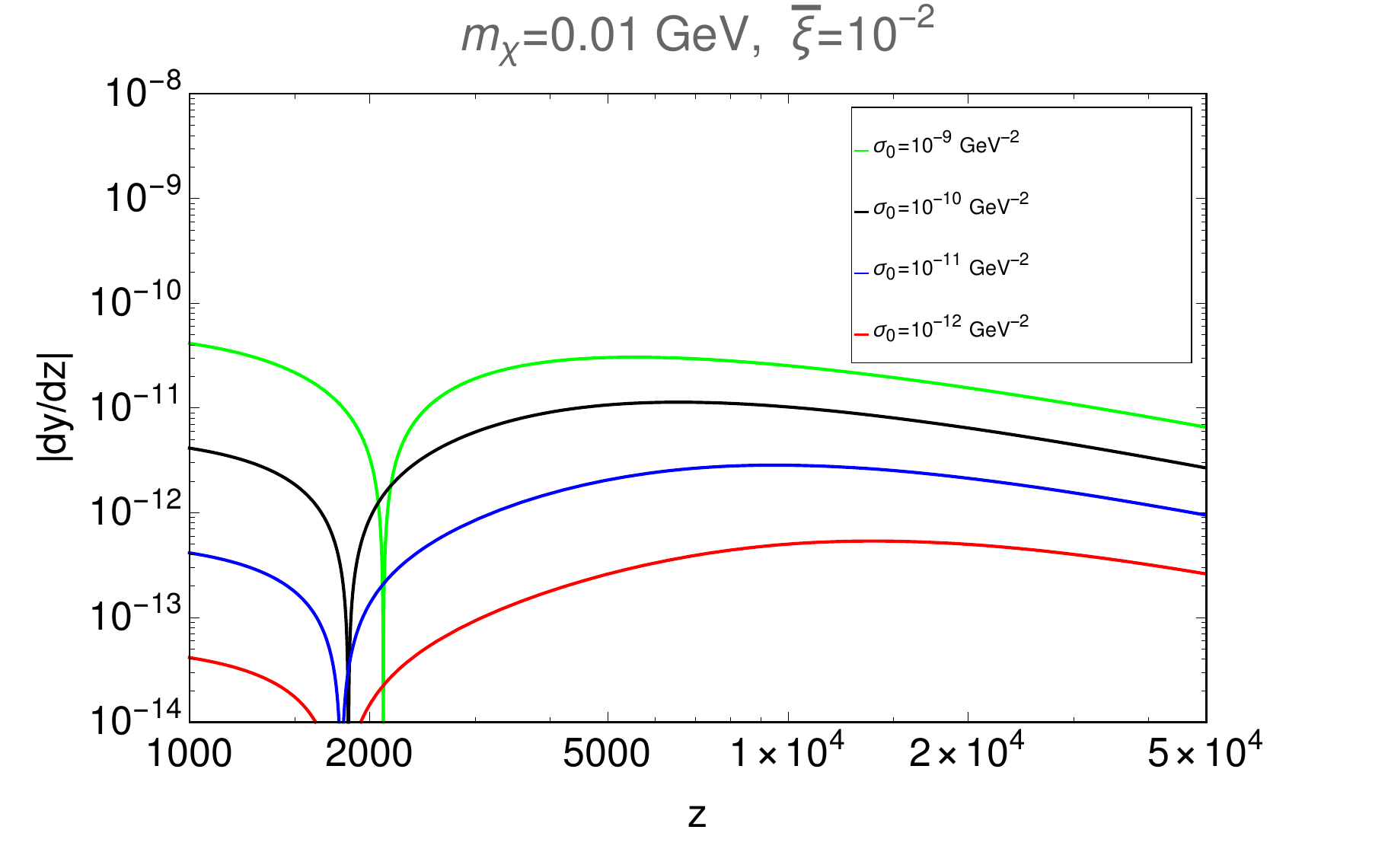}\label{p2c}}
        \subfloat[] {\includegraphics[width=3.5in,height=2.5in]{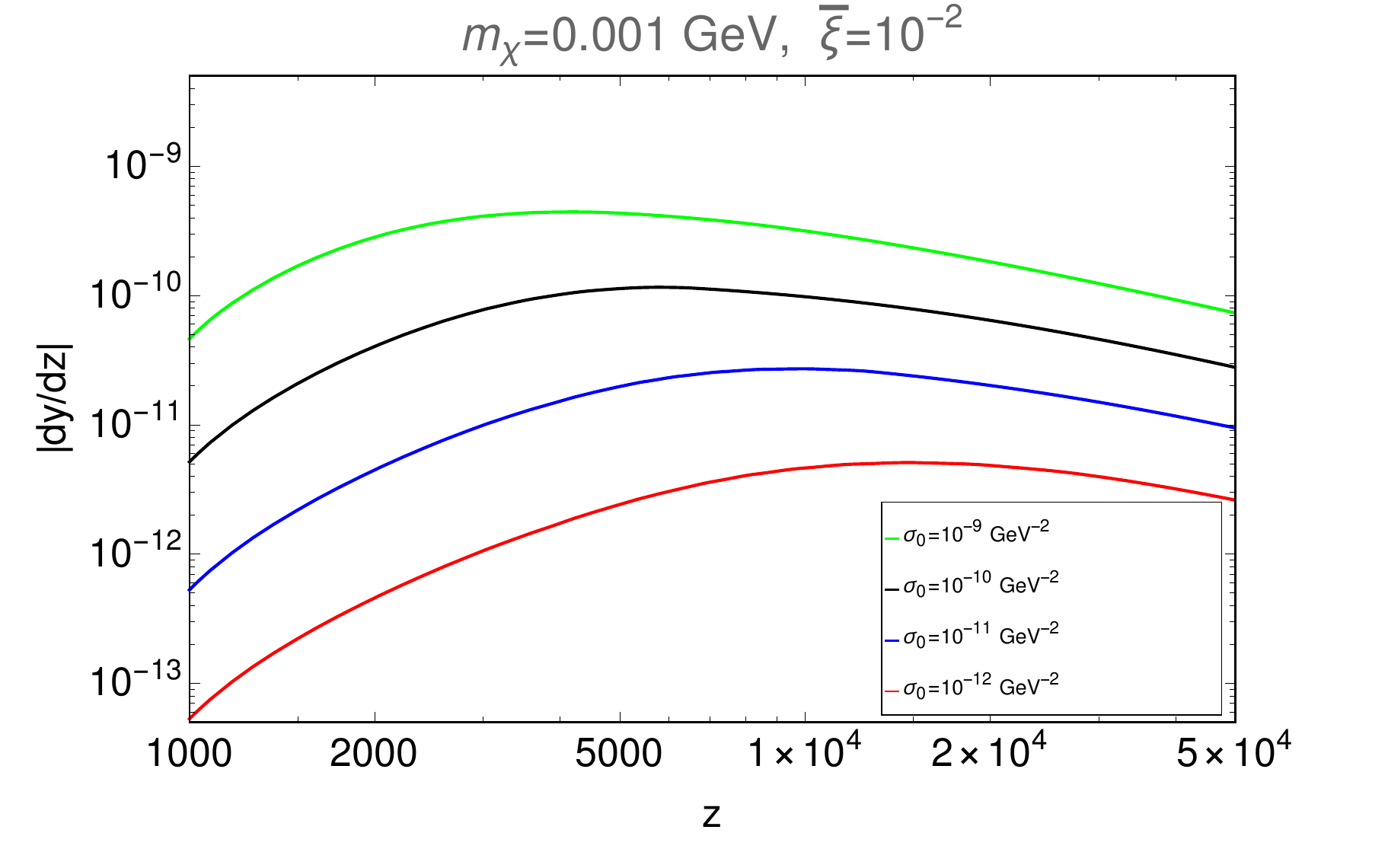}\label{p2d}}
    \end{center}
    \caption{Evolution of ${dy}/{dz}$ as a function of redshift. Here, we vary the strength of the dark matter cross-section with baryons as: $\sigma_0= 10^{-12},\ 10^{-11},\ 10^{-10},\ 10^{-9}$~GeV$^{-2}$, for fixed values of the viscosity parameters ($\gamma$ and $\zeta_0$) and mass of dark matter. In the left panels, dark matter mass is fixed to $m_{\chi}=0.01$~GeV, while in the right panels, it is fixed to $m_{\chi}=0.001$~GeV. In the figure, we have varied the value of $\zeta_0$ from $0$ to $10^{-2}$ from top to bottom.}\label{plot:2}
\end{figure*}

In figure. (\ref{plot:1}), we explore how the viscosity parameter, dark matter mass, and interaction cross-section influence each other in defining the $\mu-$type of CMB spectral distortion. From left (figure \ref{p1a} \& \ref{p1c}) to right (figure \ref{p1b} \& \ref{p1d}), we increase the dark matter mass, while from top to bottom, we increase the viscosity. In all figures, we have varied the strength of the baryon dark matter interaction cross-section from $10^{-9}$~GeV$^{-2}$ to $10^{-12}~{\rm GeV^{-2}}$. As we increase the value of $\sigma_0$, CMB spectral distortion increases. From Eq.\eqref{dmudz}, we can see that $d\mu/dz\propto dQ_b/dz$, when we increase the value of $\sigma_0$, energy transfer from baryons to dark matter increases--- Eq.\eqref{dqbdt}, and it results in a higher spectral distortion of CMB. To get the plots, we consider the initial condition for baryon temperature as $T_b=T_{\rm CMB}$, while for the dark matter temperature to be, $T_{\chi}\sim 0$ \cite{Slatyer:2018aqg}. We choose the dark matter temperature to be zero because for the considered dark matter parameter space, the dark matter decouples at a very high redshift. After decoupling, the dark matter evolves as adiabatically, $\propto (1+z)^2$, and it becomes very small compared to the baryon temperature around redshift of $\sim 2\times 10^6$. For example, for a dark matter mass of 0.01~GeV and $\sigma_0=10^{-9}~{\rm GeV^{-2}}$, the decoupling redshift for dark matter is about of $10^{11}\,$--- Eq.\eqref{axb}. Consequently, around a redshift of $2\times10^6$, the dark matter temperature is around $10^5$ times smaller than the baryon temperature. Therefore, while solving the differential Eqs.\eqref{dtbdz} and \eqref{dtxdz}, we consider dark matter temperature, $T_\chi\sim0$. In figure \eqref{p1a}, initially, the value of $|d\mu/dz|$ is large, and it starts to decrease as we go towards lower redshift. It happens because the energy transfer from baryons to dark matter, $dQ_b/dt\propto u^{-3}$. Initially, the dark matter temperature is zero, and the dominating term in $u$ is $(T_b/m_b)^{1/2}$. As we evolve the equations with redshift, the dark matter temperature starts to rise, and the term $T_\chi/m_\chi$ in $u$ becomes larger than $T_b/m_b$. Therefore, initially, $1/u^3$ starts to decrease, resulting in smaller values of $|d\mu/dz|$. After a certain point, the dark matter temperature starts to decrease due to the expansion of the Universe, causing a higher value of $1/u^3\,$--- i.e. higher spectral distortion of CMB. In addition to this $|d\mu/dz|\propto\rho_\chi/\rho_\gamma\propto(1+z)^{-1}$, thus, as we go towards lower redshift, the value of $|d\mu/dz|$ increases. In this case $|d\mu/dz|$, initially follows $T_b/m_b$ and later it follows $T_\chi/m_\chi$. The case with $\sigma_0=10^{-12}~{\rm GeV^{-2}}$ (red solid line) increases continuously. This happens because the temperature of dark matter always remains more than $10^2$ times smaller than the baryon temperature due to the small interaction cross-section between dark matter and baryons. Therefore, the term $T_b/m_b$ is always larger than the $T_\chi/m_\chi$ in $u$ for the considered redshift range. Consequently, $d\mu/dz$ follows the $T_b/m_b$ for the complete $\mu$-distortion era. As we decrease the dark matter mass, the energy transfer from baryon to dark matter increases, resulting in a higher spectral distortion of CMB--- figure \eqref{p1b}. In figures (\ref{p1c} and \ref{p1d}), we include the viscosity of dark matter. Due to the viscosity, the dark matter temperature rises. Therefore, the energy transfer from baryons to dark matter becomes less effective compared to non-viscous cases--- figures (\ref{p1a} and \ref{p1b}). This results in a smaller spectral distortion of CMB.

In figure \eqref{plot:2}, we explore the impact of viscosity and dark matter parameters on $y$-type of spectral distortion of CMB. This type of distortion comes into effect when the baryon temperature becomes $\lesssim10^5$~K, and non-relativistic Compton scattering between photons and baryons takes place. This corresponds to the redshift of about $5\times 10^4$. As we go toward lower redshifts, the value of $u$ decreases. However, it remains larger than the relative velocity between dark matter and baryons during the $\mu$-distortion era. Therefore, term $\exp(-r^2/2)$, in Eq.\eqref{dqbdt}, remains about  unity. As we go below the redshift of $\sim10^5$, $\exp(-r^2/2)$ starts to decrease. For $\sigma_0=10^{-9}~{\rm GeV^{-2}}$ and $\Bar\zeta=0$, it changes from 1 to $\sim0.8$ for a redshift range of $2\times10^6$ to $10^3$. While for $\sigma_0=10^{-12}~{\rm GeV^{-2}}$ and $\Bar\zeta=0$, it change about $10^3$ times--- from 1 to $\sim 10^{-3}$. This can be seen in figures (\ref{p2a} and \ref{p2b})--- red solid lines. In the cases with dark matter viscosity (figures \ref{p2c} and \ref{p2d}), the changes in the values of $\exp(-r^2/2)$ are not drastic compared to the previous case. For $\Bar\zeta=10^{-2}$ and $\sigma_0=10^{-9}~GeV^{-2}$: (I)  $m_\chi=10^{-2}$~GeV, $\exp(-r^2/2)$ varies between 1 and $0.5$; (II) $m_\chi=10^{-3}$~GeV, $\exp(-r^2/2)$ varies between 1 and $0.9\,$. For $\Bar\zeta=10^{-2}$ and $\sigma_0=10^{-12}~{\rm GeV^{-2}}$: (I)  $m_\chi=10^{-2}$~GeV, $\exp(-r^2/2)$ varies between 1 and $0.2$; (II) $m_\chi=10^{-3}$~GeV, $\exp(-r^2/2)$ varies between 1 and $0.6\,$.  In the presence of viscosity in dark matter, there are two sources for heating dark matter: one is the energy transfer from baryon, and the other one is the viscosity--- Eq.\eqref{dtxdz}. Therefore, the dark matter temperature may rise above the baryon temperature, resulting in a sign flip of $dy/dz$. The sign changes from positive to negative, causing a kink--- this can be seen in the figure \eqref{p2c}. As discussed earlier, by decreasing the dark matter mass, the energy transfer from baryon to dark matter increases, and we get a higher spectral distortion of CMB--- also displayed by figure \eqref{p2d}. 
\begin{figure*}
    \begin{center}
        \subfloat[] {\includegraphics[width=3.5in,height=2.3in]{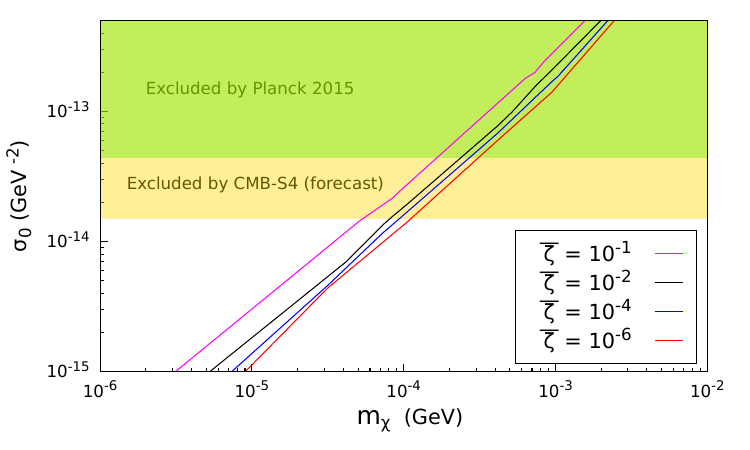}\label{p3a}} 
        \subfloat[] {\includegraphics[width=3.5in,height=2.3in]{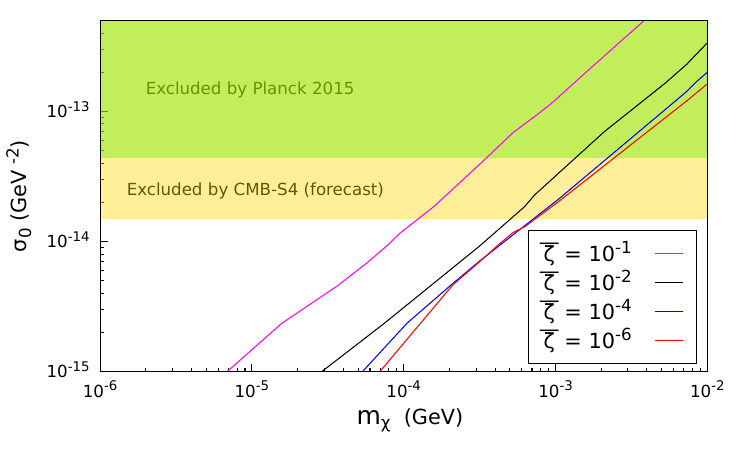}\label{p3b}}
    \end{center}
    \caption {Constraints on dark matter parameter space for $\mu$ (left panel) and $y$ (right panel) type of CMB spectral distortion for different viscosity parameters. The green shaded region corresponds to the Planck 2015 constraints (excluded) on the strength of dark matter cross-section with baryons ($\sigma_0$) and mass of the dark matter ($m_\chi$) with $95\%$ C.L., and the yellow region is forecast by CMB-S4 \cite{Boddy:2018G, PhysRevD.98.103529}. In both plots, solid coloured lines depict the upper bound on $\sigma_0$ as a function of $m_\chi$ for the corresponding viscosity of dark matter. The bounds have been obtained using PIXIE limits on $\mu$ (left figure) and $y$ (right figure) types of spectral distortion. PIXIE limit on the $\mu$-type spectral distortion is $5\times10^{-8}$; while, on the $y$-type spectral distortion is $10^{-8}$}.\label{plot:3}
\end{figure*}

In figure \eqref{plot:3}, we have constrained the dark matter viscosity using the bounds on $\sigma_0-m_\chi$ by Planck 2015 and CMB-S4 in light of $\mu$-type (figure \ref{p3a}) and $y$-type (figure \ref{p3b}) spectral distortion of CMB. The green and yellow shaded regions are excluded by Planck 2015 and CMB-S4 (forecast), respectively \cite{Boddy:2018G, PhysRevD.98.103529}. In figures \eqref{p3a} and \eqref{p3b}, the solid lines have been obtained using the PIXIE limits on $y$ and $\mu$ types of CMB spectral distortion, respectively. The PIXIE limits on $y$ and $\mu$ types of spectral distortion correspond to $<10^{-8}$ and $<5\times10^{-8}$, respectively. The coloured lines in figures (\ref{p3a} and \ref{p3b}) represent the upper bound on the dark matter baryon cross-section as a function of mass for viscous dark matter. Increasing the dark matter viscosity results in higher dark matter matter temperature. Increasing the dark matter temperature reduces the efficiency of energy transfer from baryon to dark matter. Therefore, one has to increase the interaction cross-section to transfer energy efficiently from baryon plasma to dark matter. Using this mechanism, one can constrain the viscosity of dark matter for a given parameter set. For example, for dark matter mass $\gtrsim7\times 10^{-4}$~GeV, viscosity $\Bar\zeta\gtrsim10^{-6}$ is ruled out by CMB-S4 and PIXIE constraints on $\mu$-type of CMB spectral distortion. Similarly, for  $m_\chi\gtrsim1.5\times 10^{-4}$~GeV, viscosity $\Bar\zeta\gtrsim10^{-1}$ is ruled out. In our analysis, we do not consider FIRAS limits as PIXIE limits are stringent.

\section{Conclusions}
\label{sec6}

In the $\Lambda$CDM cosmology, at redshift $z>10^6$,  the  CMB blackbody spectrum was maintained by the double Compton scattering and bremsstrahlung process.  Below redshift $z < 2\times10^6$,  bremsstrahlung and Couble Compton scattering become inefficient . If any non-standard process produces energy or photons at $z < 2\times10^6$, the CMB spectrum bears an imprint, and the blackbody spectrum cannot be restored. As a result, it establishes a Bose-Einstein spectrum distortion at redshift $5\times10^4 \lesssim z\lesssim 2\times10^6 $, and in lower redshift $10^3 \lesssim z\lesssim 5\times10^4$, it produces a \textit{y}-type distortion. The future experiment PIXIE (COBE/FIRAS) will be able to constraints on $\mu\approx5\times10^{-8}$ ($\mu\leq9\times10^{-5}$), and $y\approx10^{-8}$ ($y\leq1.5\times10^{-5}$) at $5\sigma$.


Here, we have considered the pre-recombination contribution of spectral distortion by the non-standard interaction of the baryon and the viscous dark matter. We find that for fixed dark matter mass and bulk viscosity parameter $\Bar\zeta_0$ when we increase the baryon dark matter cross section ($\sigma_0$), energy transfer from baryon to dark matter increases; hence, we see higher CMB spectral distortion (both $\mu-$ and $y-$ type). In the case of constant baryon-dark matter cross-section ($\sigma_0$) and bulk viscosity parameter $\Bar\zeta$, low dark matter mass efficiently transfers energy from baryon to dark matter resulting in higher CMB spectral distortion (both $\mu-$ and $y-$ type). Further,  for fixed dark matter mass and baryon-dark matter cross-section $\sigma_0$, when we increase the viscosity parameter $\Bar\zeta$, the temperature of the dark matter increases, and the temperature difference between baryon and dark matter decreases. Hence, the heat transfer rate $dQ_b/dt$ becomes less efficient, resulting in a smaller amplitude in the CMB spectral distortion. However, as we move towards a lower redshift, the temperature of the dark matter temperature becomes higher than the baryon temperature due to the viscous nature of dark matter. As a result,  the heat transfer rate ($dQ_b/dt$)  changes sign from positive to negative (i.e., figure \ref{p2c}), and we obtain the kink in the magnitude of $dy/dz$.

Using the PIXIE constraints on CMB spectral distortion and dark matter-baryon scattering cross-section, we have constrained the viscosity of dark matter for a given parameter set. For example, for dark matter mass $\gtrsim1.5\times 10^{-4}$~GeV, viscosity $\Bar\zeta\gtrsim10^{-1}$ is ruled out by CMB-S4 and PIXIE constraints on $\mu$-type of CMB spectral distortion. In other words, for a finite value of dark matter viscosity, the upper bound on the baryon-dark matter interaction cross-section gets modified.


\bibliography{main.bib}

\end{document}